# Convolutional neural networks for global human settlements mapping from Sentinel-2 satellite imagery

Christina Corbane[1] · Vasileios Syrris[1] · Filip Sabo[2] · Panagiotis Politis[2] · Michele Melchiorri[3] · Martino Pesaresi[1] · Pierre Soille[1] · Thomas Kemper[1]

## Abstract

Spatially consistent and up-to-date maps of human settlements are crucial for addressing policies related to urbanization and sustainability, especially in the era of an increasingly urbanized world. The availability of open and free Sentinel-2 data of the Copernicus Earth Observation program offers a new opportunity for wall-to-wall mapping of human settlements at a global scale. This paper presents a deep-learning-based framework for a fully automated extraction of built-up areas at a spatial resolution of 10 m from a global composite of Sentinel-2 imagery. A multi-neuro modeling methodology building on a simple Convolution Neural Networks architecture for pixel-wise image classification of built-up areas is developed. The core features of the proposed model are the image patch of size $5 \times 5$ pixels adequate for describing built-up areas from Sentinel-2 imagery and the lightweight topology with a total number of 1,448,578 trainable parameters and 4 2D convolutional layers and 2 flattened layers. The deployment of the model on the global Sentinel-2 image composite provides the most detailed and complete map reporting about built-up areas for reference year 2018. The validation of the results with an independent reference dataset of building footprints covering 277 sites across the world establishes the reliability of the built-up layer produced by the proposed framework and the model robustness. The results of this study contribute to cutting-edge research in the field of automated built-up areas mapping from remote sensing data and establish a new reference layer for the analysis of the spatial distribution of human settlements across the rural–urban continuum.

**Keywords** Convolutional neural networks · Remote sensing · Image segmentation · Human settlements · Built-up areas

## 1 Introduction

New ways to map and measure the built-up environment over large areas are critical to answering a wide range of research questions and to addressing policies related to urbanization and sustainability. This is particularly true in the era of an increasingly urbanized world [1]. Earth

✉ Christina Corbane
  Christina.corban@ec.europa.eu

1  European Commission, Joint Research Centre (JRC), Ispra, Italy

2  Arhs Developments S.A, 4370 Belvaux, Luxembourg

3  Engineering S.p.a, 00144 Rome, Italy

Observation (EO) has become a promising tool to provide up to date geospatial information on the status and dynamics of built-up areas and human settlements [2]. With the routine acquisition of satellite imagery and the availability of different satellite collections, several efforts have focused on mapping built-up areas at a global scale in the last decade. The most recent datasets include the Global Urban Footprint (GUF) with its 12 m product derived from TerraSAR-X imagery acquired in 2011–2013 [3]; the Global Human Settlement Layer (GHSL) with the 30 m multitemporal datasets derived from Landsat archives and showing the evolution of built-up areas in four epochs 1975, 1990, 2000 and 2014 [4, 5]; the World Settlement Footprint (WSF) with the 10 m resolution datasets based on Landsat-8 and Sentinel-1 sensors for reference year 2015 [6] and the FROM-GLC10 landcover map which includes a dedicated class for artificial surfaces derived from Sentinel-2 data acquired in 2017 [7]. Unlike the GUF which was generated from commercial imagery, all the other products were derived from free and open-access satellite image datasets, primarily from Landsat and the European Copernicus Sentinel missions. The advantages of these products are numerous and are mainly related to their free availability (absence of restrictions on their use for multiple types of applications) and most of all for the relatively low cost of their systematic update.

The methods used to produce these products and in general to extract built-up areas or artificial surfaces from remote sensing data include statistically derived indices and both supervised and unsupervised learning approaches. The first group of methods covers typically spectral indices [8–10], spectral mixture analysis [11, 12] and local/contextual image contrast/texture analysis [13, 14]. The latter includes regression analysis [15–17] and machine learning techniques, comprising mostly decision trees and random forests [18–20], support vector machines [7] and associative rule learning [4, 5].

Although some of these methods have proved to be suitable for large-area mapping of human settlements from satellite imagery, several limitations must be considered when using the information products generated from public satellite data for analytical purposes. These limitations are mostly related to accuracy, sensor-scale dependency, mapping of the extrema of the settlement density range, and the continuous monitoring of urban land cover changes. A non-exhaustive list follows below:

- High false positive and false negative error rates from the automated detection of urban land cover classes when compared to non-urban classes (e.g., bare rocks, sand dunes, bare agricultural fields, river bank lines) due to the limited actual extent of built-up areas and the discontinuous surface they compose [21];
- High disagreement on total land cover surface estimates of different sensor-derived products and high dependency on input sensor resolution of the urban land cover total estimates [22–24];
- Unsatisfactory mapping of the extrema of the settlement spatial patterns at the very low-density rural areas and the very high-density urban areas [25–27];
- Lack of a commonly approved methodology and/or a machine-based automatic and reproducible solution which allows consistent and continuous monitoring of global urban land cover changes across time and across different sensors [2, 28, 29].

Compelling challenges and opportunities still lie ahead in high-resolution mapping and accurate classification of built-up areas over large areas. A key issue in this context is up-to-date and reliable information on the status and development of the human settlements. The availability of free and open remotely sensed big data streams has brought significant innovations in the field of automatic information extraction from satellite imagery. There is an increasing need to mine the large amount of earth observation data delivered in a free and open way by some of the new generation of satellites, especially the Sentinel missions. Operational since 2017, the Sentinel-2 mission of the European Copernicus program provides a 5-day repeat cycle and a span of 13 spectral bands at a spatial resolution as high as 10 m. Sentinel-2 has great potential for mapping and monitoring built-up areas on a global scale [7, 30, 31]. Novel approaches for mapping human settlements are needed to deal with the increased spatial and temporal resolution of Sentinel-2.

## 1.1 Background

Advances in deep learning (DL) have led to leaps in the fields of computer vision, speech recognition and natural language processing.

Whereas the task of built-up areas extraction from remote sensing data has a number of unique challenges, primarily related to the sensor and the features to be detected, it draws concepts and theories from computer vision, signal processing, statistics and machine learning [38]. Since 2014, the remote-sensing community has shifted its attention to DL for addressing different application domains [32] such as image registration [33], image fusion [34, 35], change detection [36] and object detection [37]. However, image classification is the remote sensing field where DL has gained most of its popularity [32].

Recent applications in remote sensing have used DL approaches for image classification tasks at which the purpose was the labeling of single pixels or regions of an image according to two or more classes [39–41]. DL methods have experimentally proved to outperform state-of-the-art machine learning methods (e.g., Support Vector Machines, Random Forests) [42] for the classification of both optical (hyperspectral and multispectral imagery) [41, 43], radar imagery [44], change detection [45] and for the extraction of different land cover types such as roads [46], crop types [40] and buildings [47].

Ball et al. (2017) [38] provide a comprehensive survey of image classification works in remote sensing that rely on DL approaches, while the review paper of Ma et al. (2019) [42] on DL approaches covers nearly every application and technology in the field of remote sensing, ranging from preprocessing to image fusion, object detection and land cover mapping. A recent study suggested that deep learning is suitable for capturing the fine features of complex urban areas and performs better than conventional threshold-based methods, traditional supervised classifications and machine learning approaches [48]. In particular, architectures building on convolutional neural networks (CNNs) have become viable solutions for remote sensing image classification where traditional handcrafted feature engineering and domain-knowledge methods fail because of the limited generalization capabilities of the algorithms, the inter-class similarity, the intra-class variability as well as the changing image acquisition conditions [49, 50].

Differently from other DL approaches, deep CNNs were specifically designed for image classification; nevertheless, they can be easily adapted to solve image segmentation problems by performing pixel-wise classification [51]. The hierarchical features of the input image data are modeled naturally by the CNN hierarchical structure, a fact that boosts the CNN performance in satellite image classification in general and facilitates the extraction of built-up features in particular. Another main advantage of CNN architectures over other established methods used for generating the global maps of built-up areas is their capacity to be integrated with mature frameworks of image pre-processing and standardization tools providing shift-invariant and contrast-invariant image local transforms [52].

Recognizing the inherent advantages of convolution operations in the characterization of the built-up environment in remote sensing data, a significant amount of works have recently explored the potential of diverse CNN architectures for mapping built-up areas from different types of sensors and different spatial resolutions: Synthetic Aperture Radar [44], high and very high spatial resolution imagery [48, 53, 54] and aerial imagery [55] (i.e., with a ground sampling distance equal to or even less than 1 m). However, little effort has been directed toward the challenge of large-scale built-up areas mapping with CNN from data of lower spatial resolution such as the ones powered by Sentinel-2. The works of [56, 57] represent a significant advancement in that direction. In particular, the framework of human settlements mapping proposed at 20 m by [57] is a step-forward toward a global scale model. Despite the demonstrated generalization and upscaling capabilities of their proposed framework, the authors failed to implement the CNN model in rural areas, which represent one of the main challenges in built-up areas mapping from satellite data at global scale.

## 1.2 Challenges addressed in this work

When deploying CNNs on large geographical areas or at global scale, four main issues should be taken into consideration:

- The necessity to develop a model flexible enough to be applied to a global carpet of satellite data entailing the design of a sound training approach, a strategy for transfer learning and a plan for the consistency verification of the classification output.

- The substantial amount of training data required for training complex models. In the case of built-up classification, the training samples should cover different building types (e.g., residential and industrial buildings of different sizes, colors and rotations) in various types of landscapes (e.g., dense urban areas, rural areas, desert landscapes, built-up areas mixed with neighborhood green spaces);

- The increased need for computational processing resources, especially for adjusting and fine-tuning multiple and/or complex models;

- The requirement for CNN architectures that are robust to noise in satellite imagery (e.g., presence of snow, clouds, haze) and to other seasonal effects. This feature would enable the generalization capacity of the models over large areas and the extraction of built-up areas with comparable efficacy along the urban–rural continuum.

In this work, we propose a Neural Computing framework tailored for global scale mapping human settlements at a spatial resolution of 10 m, from a cloud-free composite of Sentinel-2 data for reference year 2018. The output is a global map of built-up areas expressed in terms of a probability grid.

The main contributions of the work can be summarized as follows:

- A new framework for pixel-wise large-scale classification of built-up areas from a Sentinel-2 image composite at a spatial resolution of 10 m has been developed, named GHS-S2Net (GHS stands for Global Human Settlements, S2 refers to the Sentinel-2 satellite) (Sect. 2.3);

- A multi-neuro modeling methodology is proposed following the Universal Transverse Mercator (UTM) grid zones schema and a systematic two-stage sampling within each UTM grid zone (Sect. 2.3.1);

- Transfer learning is implemented following two separate approaches depending on the availability of reliable training data at the different UTM zones: a close range transfer learning within each UTM grid zone and a far range transfer learning from one UTM grid zone to neighboring data-poor zones (Sect. 2.3.3). In this work, transfer learning does not obey the most dominant definition of using the weight values of pre-trained models from different domains. As a concept herein, it is closer to the verification of the generalization capacity of the models when the training and testing data do not necessarily follow similar statistical distributions;

- An extensive assessment of the models output, that is based on an independent validation using fine-scale digital cartographic reference data reporting the footprint of every single building for 277 sites around the globe (Sect. 3.4).

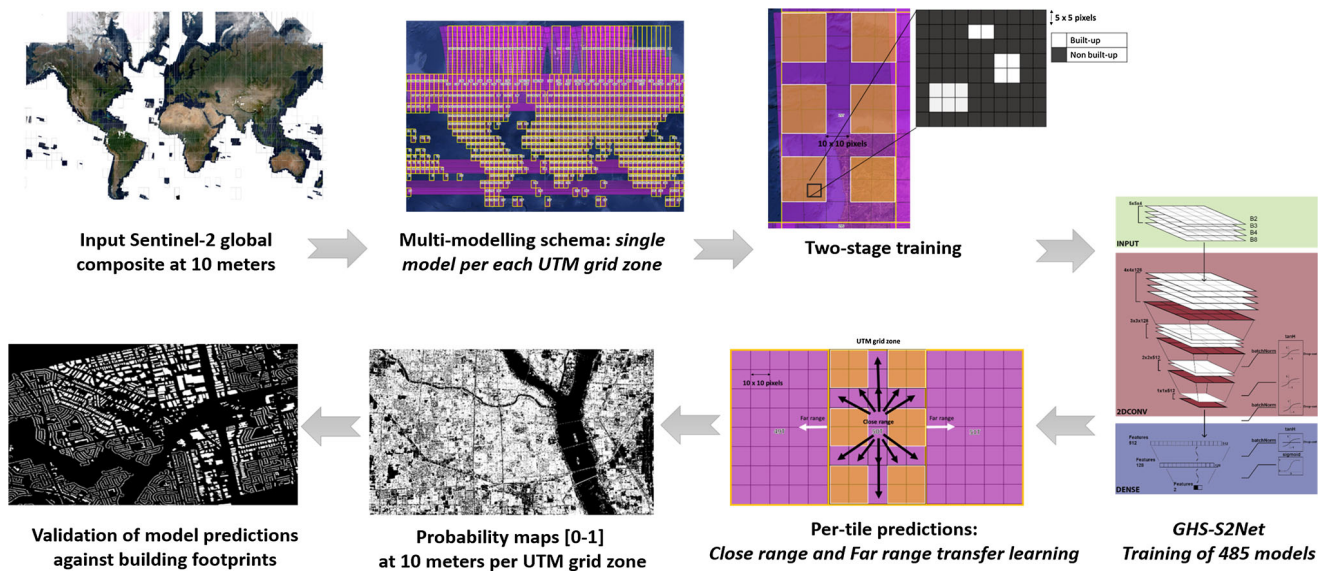

**Fig. 1** Overview of the proposed framework for global human settlements mapping from Sentinel-2 image composite

The proposed framework for built-up areas classification at global scale is summarized in the illustrative diagram of Fig. 1. Details on each of the processing steps are provided in the subsequent sections.

The new framework leverages the JRC Big Data Platform (JEODPP) [58] for the storage of the global input data and for optimized fast parallel processing using the high-performance Graphics Processing Units (GPUs). This dedicated infrastructure allows tackling the challenges of large-scale processing, boosting the CNN training, and enhancing the prediction accuracy through duly fine-tuning of the models.

## 2 Input data and methods

### 2.1 Sentinel-2 cloud-free image composite

The input data for human settlements mapping over the entirety of the landmass (excluding Antarctica) consist in a global cloud-free image composite for reference year 2018 derived from Sentinel-2 satellite data of the European Copernicus Earth observation program. Sentinel-2 mission offers a great potential for fine-scale mapping and monitoring of built-up areas thanks to high spatial and temporal resolutions, with a five-day revisit time and decametric resolution [31]. However, the selection of the best available scenes, their download from the dedicated data hubs together with the requirements in terms of storage and computing resources pose restrictions for large-scale mapping. Pixel-based compositing is an approach to leverage the large volumes of available data, while effectively mitigating cloud and aerosol contamination as well as data gaps in the archive [59]. This method has been recognized for being a valuable tool for large area applications using high spatial resolution optical data [60]. Accordingly, the image composite was generated in and exported from Google Earth Engine [61]. The methodology used for the selection of the satellite imagery and for image compositing is based on a data-driven approach which uses a summary statistic for aggregating the pixel time series (i.e., the 25th percentile). A detailed description of the workflow is presented in [62]. The output image composite consists of a global scale raster grid of four spectral bands derived from top of atmosphere Sentinel-2 image tiles (B2: Blue, B3: Green, B4: Red and B8: Visible and Near Infrared) with a spatial resolution of 10 m. It was produced and tiled following the UTM system with each tile having the projection of the UTM zone (UTM/WGS84 projection) to which it corresponds to. There are in total 615 grid zones with data covering mostly mainland and islands (Fig. 2). The full dataset has a total volume of 15 TB and is hosted on the Big Data platform of the Joint Research Centre (JEODPP). The raster data have been stored in 16-bit geotiff format. The dataset can be freely accessed and downloaded from the Open Data Catalogue of the Joint Research Centre of the European Commission[1] [63].

---

[1] https://data.jrc.ec.europa.eu/dataset/0bd1dfab-e311-4046-8911-c54a8750df79.

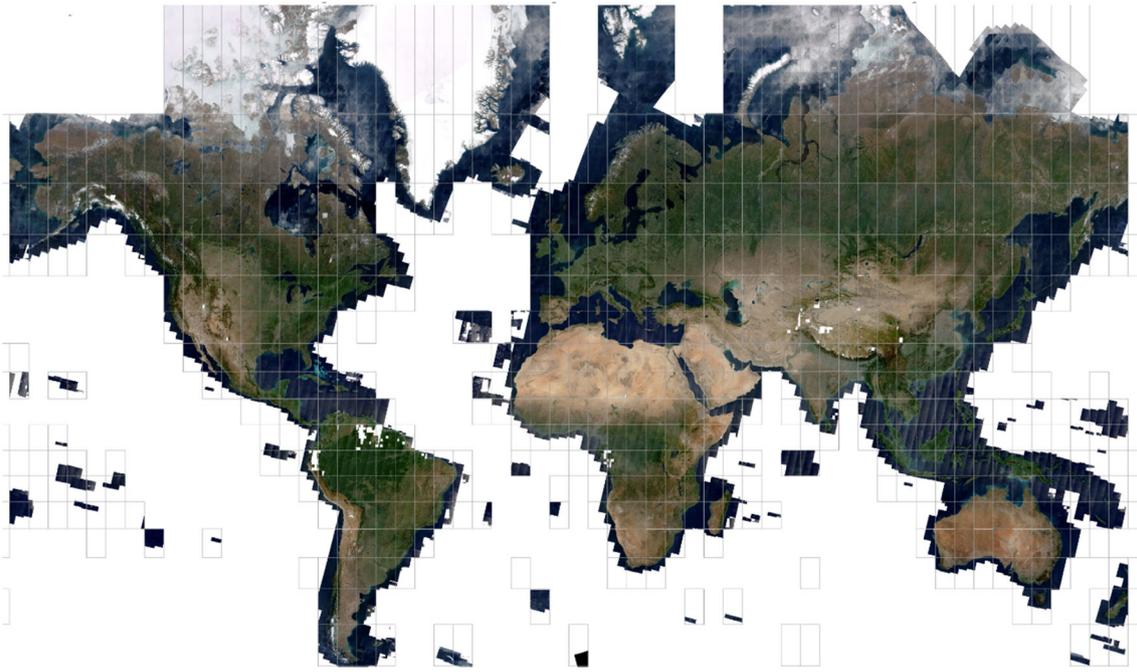

**Fig. 2** Overview of the cloud-free Sentinel-2 image composite organized by UTM grid zones

## 2.2 Model input data: learning sets

A sensitive point regarding CNNs is the amount of training data required to properly adjust the network parameters. A large source of free and open access datasets describing built-up areas was collected with different levels of details, completeness, consistency and accuracy. Since the aim is to achieve a stable and at the same time detailed and accurate delineation of built-up areas, the most detailed datasets describing built-up areas were compiled from public sources: The Global Human Settlement Layer (GHSL_BU), the European Settlement Map (ESM_BU), the Facebook high resolution settlement data (FB_HRS) and the Microsoft building footprints (MS_BFP) described hereafter.

### 2.2.1 Global Human Settlement Layer built-up areas

GHSL_BU was derived from automatic classification of Landsat 30 m-resolution data of the year 2014 as described in [5]. The method for mapping built-up areas from Landsat data at global scale builds on the Symbolic Machine Learning (SML) classifier which automatically generates inferential rules linking the image data to available high-abstraction semantic layers used as training sets [64]. The product is provided with a spatial resolution of 30 m. Despite the overall good performance in depicting built-up areas at global scale, the GHSL_BU suffers from under-detection problems in sparsely built-up areas and mainly in rural African landscapes.

### 2.2.2 European Settlement Map

ESM_BU is the 2 m resolution land cover class "built-up area" produced by the automatic classification of the Copernicus VHR_IMAGE_2015 collection which covers 39 European countries (EEA39) with various earth observation sensors. The built-up areas extraction has been achieved through supervised learning with the SML classifier along with textural and morphological features [65]. The ESM_BU is currently the most detailed map of built-up areas available for Europe. The main issue in this layer is the presence of false alarms, in particular over mountainous areas and sand beaches as well as the absence of cloud-free satellite data in some regions resulting in large data gaps observed in certain urban areas (e.g., UK (Manchester, Peterborough, Reading, Luton, Coventry) and Ireland (Dungarvan)).

### 2.2.3 Facebook high-resolution settlement data

The FB_HRS data used in the study are derived from the high-resolution settlement grids produced by Facebook [66]. The dataset was made available for public use in the frame of "Data for Good" Facebook program that supports international humanitarian efforts [67]. The settlement areas of FB_HRS were automatically delineated by a CNN classifier working over sub-meter resolution optical satellite imagery and using fine-scale open-source training data as Open Street Map (OSM) [68]. The 30 m spatial resolution derived data [67] have been used in the present study. At the time we compiled the FB_HRS data, 150 countries were covered by the FB_HRS including large parts of South America, Africa, Europe and Asia. According to the information available on a subset of 194 countries, the image data supporting the FB_HRS spatial delineation were collected in the time range from 2002 to 2017, with a temporal surface-weighted average centered in the year 2013. Based on our internal quality control procedure, the precision of these data was particularly remarkable in rural areas flagging (at a spatial resolution of 30 m) the presence of single isolated houses and small rural hamlets precisely. Commission errors were noticed occasionally in rural areas, especially in correspondence with dense forest patterns. The mapping of large urban areas as accounted by the FB_HRS data turned to be more problematic; in these areas, remarkable systematic omission errors were noticed.

### 2.2.4 Microsoft building footprint data

The MS_BFP is a vector data derived from the work of the Microsoft map team and available for public use in the OSM community. The data were automatically extracted by the Open Source CNTK Unified Toolkit developed by Microsoft. CNTK and the ResNet34 with RefineNet up-sampling layers were applied to detect building footprints from the Bing imagery that may include VHR satellite and airborne sensors [69]. The MS_BFP data were made available in vector format at a nominal scale of 1:10.000, thus supporting a detailed rasterization at $1 \times 1$ m of spatial resolution successively aggregated to $10 \times 10$ m resolution used in this study. At the time we compiled the MS_BFP data, information about four countries was available: United States, Canada, Uganda and Tanzania. Despite the detailed representation of single buildings, the MS_BFP data suffer from omission errors referring to large industrial buildings and fewer errors related to over-detections of buildings in mountainous and agricultural areas.

Table 1 gives an overview of the specific training sets used for adjusting the models with respect to the following characteristics: spatial resolution, coverage, source image collection date used for layer production, identified issues as well as the number of pixels (total and relative percentages) used as training samples. Figure 3 displays the selected information sources for training the models by geographic area.

Due to the overall quality and spatial detail of the training data and to the variability in both the spatial coverage and the type of issues associated with each dataset, a hierarchical process was implemented for selecting the best data available locally: the priority was given first to MS_BFP and ESM_BU which are the closest proxies to the built-up areas to be derived from 10 m resolution satellite data. They were followed by the FB_HRS and finally by the GHSL_BU, which is the least detailed representation of built-up areas.

### 2.3 *GHS-S2Net* building blocks

The purpose of the proposed CNN model named here *GHS-S2Net* is to perform pixel-wise classification of built-up areas at a spatial resolution of 10 m. The concept of "built-up area" applied here is consistent with the definition adopted in the framework of GHSL which is "the union of all the satellite data samples that corresponds to a roofed construction above ground which is intended or used for the shelter of humans, animals, things, the production of economic goods or the delivery of services" [70].

Pixel-wise grouping is equivalent to the standard image segmentation process, i.e., partitioning of the image into multiple segments corresponding to individual pixels or homogenous areas. *GHS-S2Net* architecture builds on the CNN configurations described in [71]. A schematic representation of the *GHS-S2Net* is visualized in Fig. 4. The two major drivers that framed the design of this CNN model are explained below:

- Firstly, given that the target to be recognized ranges in size from single residences until block of contiguous buildings, the model capacity should allow the collection and distillation of the fine information provided by either the single pixels or the small sized groups of pixels consisting of homogeneous characteristics. Unlike popular tasks for natural image segmentation and object localization where there exist sizeable image regions with common characteristics (color, texture, connectivity, etc.), the size of the objects to be recognized herein varies from 10 m (the finest resolution associated with a single pixel) to some dozens of meters. Consequently, the contextual information that surrounds one pixel and accommodates the prominent features can be expressed by narrow image windows (patches) having a size of few pixels. An extensive experimentation specifically for Sentinel-2 imagery with respect to the optimal size of an image patch at which the convolution performs efficiently is presented in [71]. In the present study, an image patch of size $5 \times 5$ has been selected as input image to the CNN, whereas the convolution of the image is achieved through successive kernels of size $2 \times 2$ with stride $1 \times 1$. At this narrow representation and with the intention of avoiding losing essential information, no pooling layers have been employed to reduce further the spatial size.
- Secondly, the motivation was to design a lightweight model that could serve adequately the chosen multi-

**Table 1** Summary characteristics of the training sets

| Training set | Pixel size (m) | Coverage | Time stamp | Advantages | Constraints | BU samples (resampled at 10 m) | |
|---|---|---|---|---|---|---|---|
| | | | | | | Number of pixels | % |
| GHSL_BU | 30 | Global | 2014 | Complete global coverage | Lower spatial resolution than the data under processing, thus including relatively higher error rates | 1.49E+09 | 28.29 |
| ESM_BU | 2 | European | 2015 | High precision from very higher resolution input data | Limited geographical coverage, large no data zones over some cities | 5.31E+08 | 10.04 |
| FB_HRS | ∼ 30 | 194 countries | 2002–2017 | High precision derived by aggregation of very higher resolution input data | Limited geographical availability, systematic false negative in dense urban areas, sporadic false positives | 2.59E+09 | 49.06 |
| MS_BFP | vector (rasterized at 1 m and aggregated to 10 m) | 4 countries | – | High precision with delineation of single buildings from very high resolution input data | Limited geographical availability, sporadic false negative in industrial areas, sporadic false positives in specific landscapes (Canadian lakes, mountainous areas), unknown imagery date | 6.66E+08 | 12.61 |

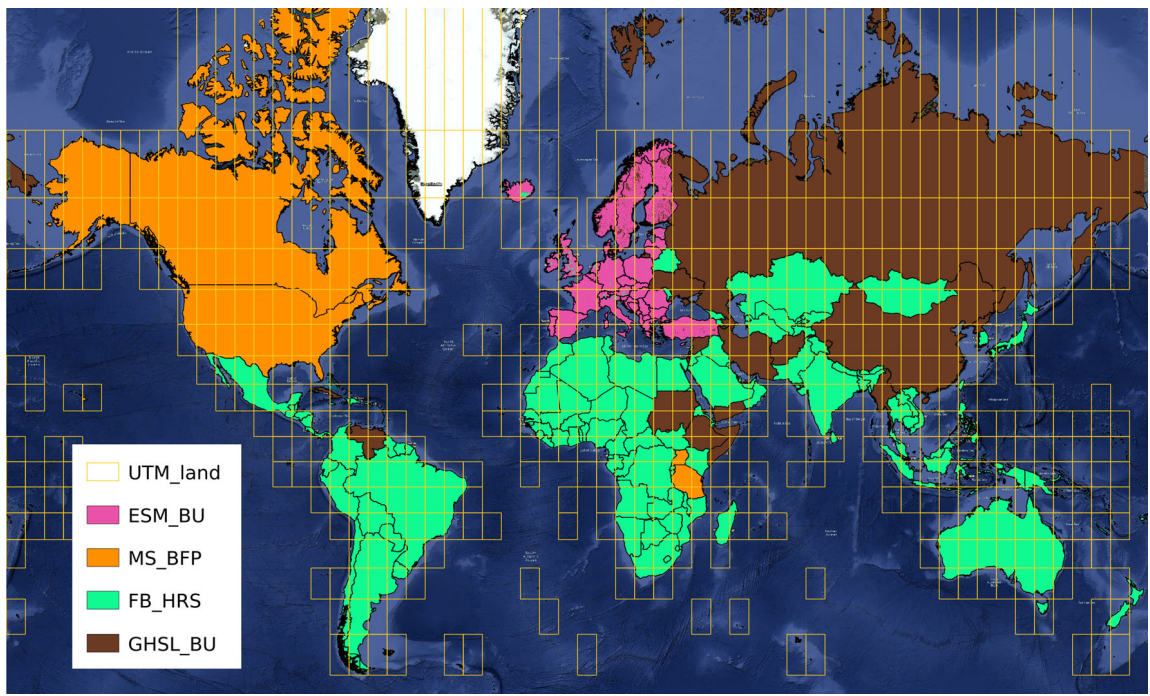

**Fig. 3** Spatial distribution of the training datasets at country level. The colors correspond to the valid data masks of the learning sets over land (color figure online)

modeling approach and allow several degrees of flexibility in terms of distributed computing. The total number of model parameters is 1,448,578 (1,447,042 trainable and 1,536 non-trainable), 95 times less than VGGNet [72] and 2.7 times less than GoogleNet [73]

(indicative CNNs). While the number of 2D convolutional layers is limited to 4 layers and the number of flattened layers to 2, the number of parameters has been increased due to the high number of filters. Tests showed that the specific CNN topology can perform quite well even if the number of filters is smaller, yet we decided to keep the number of filters high in order for the model to capture very subtle details. This lightweight topology facilitates the algorithm execution across heterogeneous GPU modules throughout the prototyping and operational phase. Additionally, it enables smoothly the multi-modeling deployment at which a different model has been trained over every UTM zone, capturing more precisely the local characteristics and the variance along similar geographical regions.

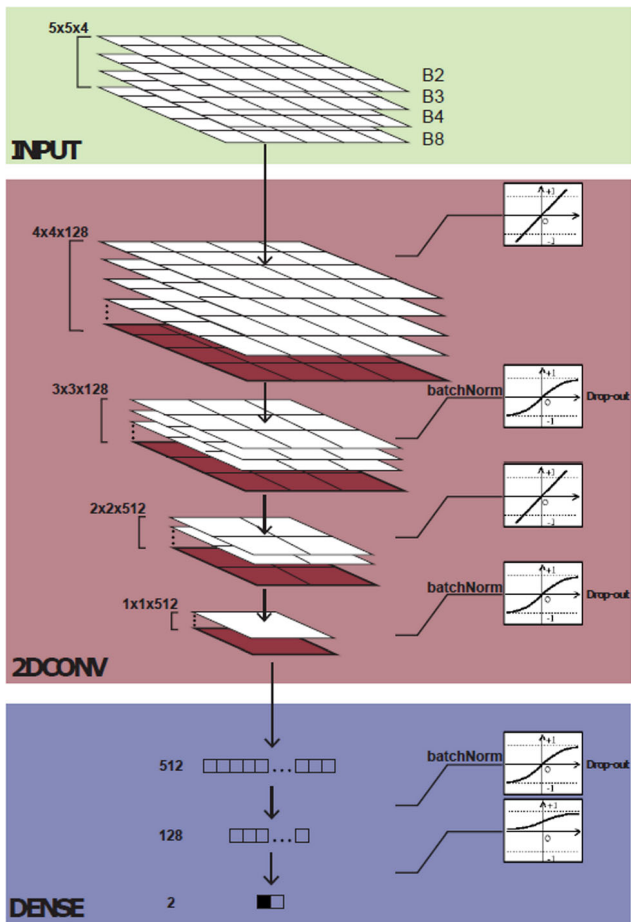

**Fig. 4** Schematic representation of the layers that compose the GHS-S2Net architecture

The *2DCONV* block as shown in Fig. 4 comprises two successive stacks where 2 × 2 convolution takes place and the linear and the hyperbolic tangent activation functions (*tanh*), respectively, transform the signals across the network layers. Although the rectifier activation function and its variants have been used widely in the various deep neural network architectures due to their robustness against the vanishing gradient problem [74], our experimentation indicated that by using a smaller number of neural network layers, the functional mapping via tanh activations captures better the complexity of the features with respect to the Sentinel-2 imagery. Besides, the tanh function is more suitable in the case of optimization with stochastic gradient descent where sigmoid function shows sharp damp gradients during backpropagation as well as gradient saturation [75]. The alternation with linear mappings results in a cost-effective solution in terms of computations. Speed-up of the training process and remedy to the effect of the internal covariate shift is provided through data batch normalization operations [76]; at each data batch, transformation is performed by keeping mean activation close to 0 and the activation standard deviation close to 1. A subsequent dropout regularization layer [77] has been used to prevent overfitting, with a ratio of 0.1 of neurons not considering at each update during the training phase.

The sigmoid function has been employed only for the last layer and maps the model output into the range [0,1], giving rise to the probability of a pixel to belong to the class built-up.

### 2.3.1 Two-stage training approach

We propose a two-stage training approach at which a single model per each UTM grid zone has been trained in accordance with the zones used for generating the Sentinel-2 image composite. This multi-modeling approach aims at capturing the variations in the Sentinel-2 data and the diverse characteristics of human settlements (in terms of size, shape, morphology and structure). Furthermore, rather than training a very complex single model that would need big volumes of representative data, the training of several relatively light CNNs facilitates the modeling of local features and distributes effectively the computational load into several machines by increasing significantly the total throughput. Each UTM grid zone covers an average area of 447,650 km$^2$ (area calculated in equal area projection). This type of data splitting is prone to containing various types of built-up areas and settlement patterns across heterogeneous landscapes even within the same UTM zone. Besides, the semantic classes of "built-up" and "non built-up" are unevenly distributed spatially and their frequencies are highly varying. The class "built-up" is very rare compared to the non-built-up class (See Supplementary material R1) (2% of the training samples (5 × 5 pixel blocks) represent built-up, while 98% represent non-built-up). To tackle this uneven distribution of training samples, each UTM grid zone was split into tiles of 100 × 100 km$^2$, which is consistent with the tile size of the Sentinel-2 granules (purple cells in Fig. 5). The two stages are described below:

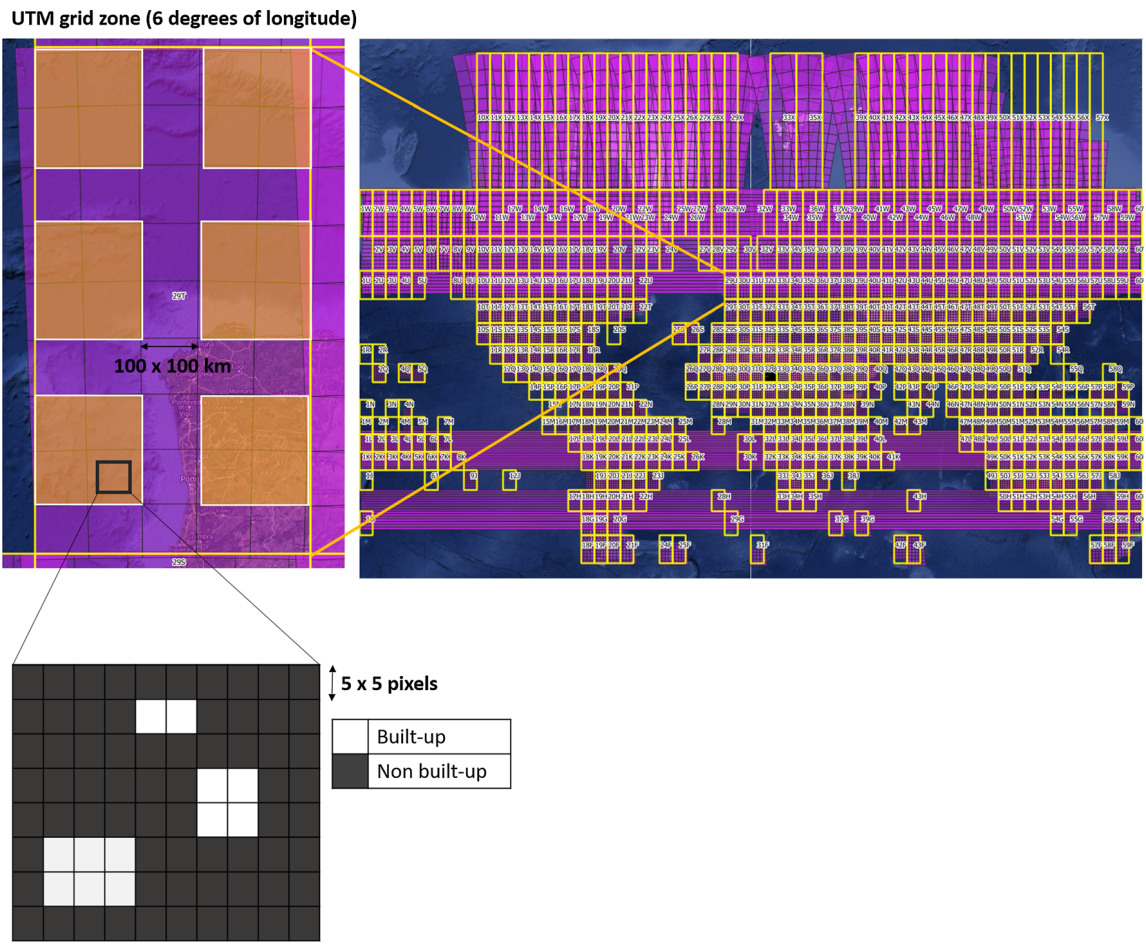

**Fig. 5** Two-stage training in which one model is trained per UTM grid zone (color figure online)

(1)  select systematically 50% of the $\sim 100 \times 100 \ \text{km}^2$ tiles of the UTM grid zones for the model training (orange boxes in Fig. 5);

(2)  consider all built-up patches ($5 \times 5$ blocks of pixels of 10 m containing at least one built-up pixel) falling within the selected $100 \times 100 \ \text{km}^2$ tiles and randomly sample 60% of the non-built-up patches uniformly with respect to their frequency in the tile (checkerboard in Fig. 5). The training of the models per UTM grid zones is done by grouping the built-up and non-built-up patches into mini-batches of 200,000 samples (where the steps per epoch depend on the training size of each UTM zone) as a compromise between computational constraints and the need to converge to a global optimum. A special attention is given to UTM grid zones largely covered by water surfaces and no data in the Sentinel-2 image composite. In such cases, all the tiles of the valid data domain are considered in the training phase without applying any sampling approach.

### 2.3.2 Per-tile predictions

As described previously, the CNN model consists of encoding layers solely, through which the information existing into image blocks of size 5 rows $\times$ 5 columns $\times$ 4 bands is multiplexed and transformed to a single value, denoting the probability of the central pixel of the $5 \times 5$ block to belong to the built-up class. The prediction phase has been performed with tiles of size 10,000 rows $\times$ 10,000 columns $\times$ 4 bands. A sliding window of size $5 \times 5$ pixels has been applied to produce the 5 rows $\times$ 5 columns $\times$ 4 bands input blocks. Constant-value image padding has been also implemented in order for the pixels at the image border to be correctly inserted into the $5 \times 5 \times 4$ input blocks. The predictions of the model are given in vector format having exactly the same size as the rows and columns product of the original input tile.

### 2.3.3 Close range and far range transfer learning

Transfer learning is a paradigm in DL to solve a target problem by reusing the learning with minor modifications from a different but related source problem. Qin et al. [78] review transfer learning in remote sensing applications and categorize the methods into four families depending on what is being transferred:

- instance-based transfer which uses partial training samples in the source domain to improve the performance of the model of the target domain [79];
- feature representation-based transfer [80] which assists the target domain classifier to learn a more effective feature expression from the source domain and improve its performance;
- relational knowledge transfer [81] where knowledge among the data in the source domain is transferred to the target domain;
- parameter-based transfer [82] considers that the source domain classifier and target domain classifier have the same optimal parameters, which can be found from the source domain classifier and then used for the target domain classifier.

Another more general classification of transfer learning methods considers the availability of labeled data and categorizes the methods into three sub-settings [83]: inductive transfer learning, when labeled data in the target domain are available; transductive transfer learning, when solely labeled data in the source domain are available; and unsupervised transfer learning, when labeled data do not exist in either the source or target domain.

One of the goals of this work is to address the following aspects of the pixel-wise classification: the computation time for training a big number of models for every UTM grid zone and the availability and precision of the training data. Parameter-based transfer learning was adopted in a transductive transfer learning framework tailored to the training strategy described in the Sect. 2.3.1. This includes a close range and a far range transductive transfer of model parameters (Fig. 6):

- The close range transfer learning consists in training the model with a subset of the input data in a given UTM grid zone (following the method described in Sect. 2.3.1) and applying it to all the $100 \times 100 \text{ km}^2$ tiles falling within the same UTM grid zone. This approach allows speeding up the training process of 485 different models and producing the predictions of a total of 30,000 tiles. It also helps overcoming overfitting issues;
- The far range transfer learning consists in training the model with detailed samples such as MS_BFP and FB_HRS in a given UTM grid zone and applying it to a neighboring zone or to zones with similar landscape and built-up typology, at which labeled samples are scarce or zones where only GHSL_BU training datasets are available. This approach allows refining the predictions and testing the generalization capabilities of the *GHS-S2Net* model.

## 2.4 Processing infrastructure

The computing-intensive workflow was executed on the JEODPP infrastructure. The JEODPP is a versatile platform with multi-petabyte scale storage (14 PiB currently) co-located with computational capabilities [58]. The platform is based on commodity hardware and open-source software stack including the EOS storage technology developed by the European Organization for Nuclear Research (CERN) [84]. The platform has been recently upgraded with a series of GPU nodes to speed-up machine/deep learning applications. Currently, there are 5 GPU nodes equipped with different types of GPU modules and memory per module. For the training of the *GHS-S2Net* models, as well as for the prediction phase, 2 GPU nodes were used: the first with 4 Quadro RTX 6000 with 24.2 GB of memory and the second with 2 T V100-PCIE with 32.5 GB of memory. Dedicated Docker images integrating the necessary deep learning packages were created to run all the experiments.

**Fig. 6** Example of close range and far range transfer learning according to the two-stage training approach. Close range transfer learning is performed in this illustrative example within UTM grid zone 50 T and far range transfer learning is done by transferring the model parameters from UTM grid zone 50 T to nearby zones 49 T and 51 T

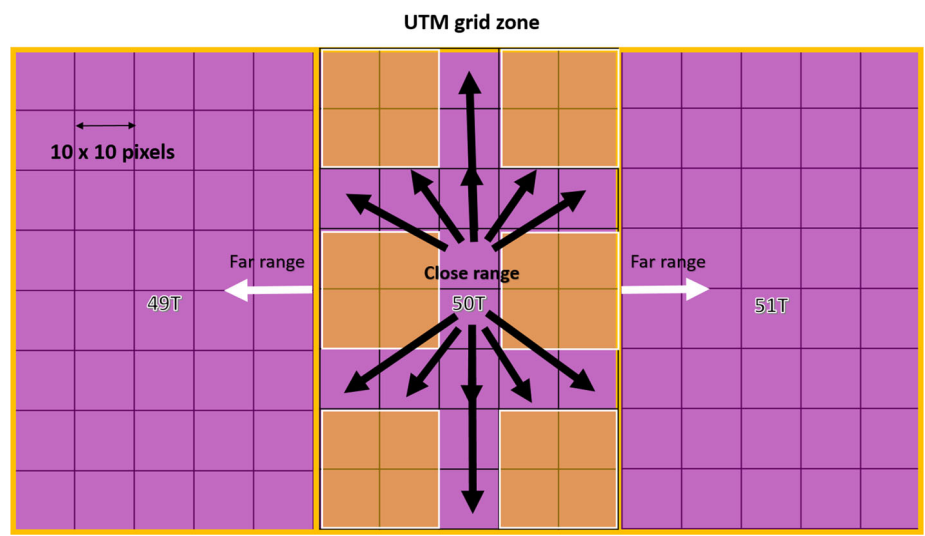

# 3 Results

## 3.1 Training phase of CNN models per UTM grid zone

### 3.1.1 Hyper-parameters tuning

During the training phase of the model per each UTM grid zone, 10% of the training data were reserved for validation in order for the CNNs to prevent over-fitting. The input Sentinel-2 composite data were rescaled in the range [0,1]. The number of epochs to train the models was set to 25 iterations. The weights were initialized based on uniform distribution with bounds [−0.1065, 0.1065]. Finally, the Adam stochastic optimization with a learning rate of 0.0001 has been used to optimize the binary *cross-entropy*, log loss function:

$$L(\mathrm{y}, \hat{y}) = -\frac{1}{N} \sum_{n=1}^{N} [y_n \log \hat{y}_n + (1 - y_n)(1 - \log \hat{y}_n)] \qquad (1)$$

where $N$ is the number of training samples, $y$ is the vector of the real target values of the training set in binary coding, and $\hat{y}$ is the vector of the model responses in the continuous range [0, 1]. The *cross-entropy* loss has fast convergence rate and is numerically stable when coupled with *sigmoid* normalization [85].

### 3.1.2 Performance evaluation

For evaluating the classification performance of the models during the training and prevent overfitting, a fraction representing 10% of the training data was used for validation. Figure 7 shows the progress of the average loss curves produced by 485 *GHS-S2Net* models during their training and validation which last 25 epochs. Every model corresponds to one UTM grid zone, resulting in 485 out of 615 grid zones that refer to landmass with presence of built-up according to the learning sets. The learning curves show that both the average training loss (green curve) and validation loss (red curve) decrease rapidly to a point of stability with a convergence around 12 epochs. The fact that the gap between the two curves is very small even for the first 5 iterations and that it completely disappears around

12 iterations after, shows that the size of the training sets, selected following the two-stage training approach, is optimal and that the models have good generalization capacity [86].

## 3.2 Computational performance of the GHS-S2Net models during the training and prediction phases

Both training and prediction were performed on GPUs and their runtime is reported in Fig. 8. The reported elapsed time refers to every UTM grid zone predominantly covered by land (204 grid zones) and those zones predominantly covered by water (281 grid zones). In inland tiles, more training samples are usually fed to the *GHS-S2Net*, while in water tiles the number of training samples is smaller. The stacked bar plots show that the average training time is around 3600 s, while the prediction time is around 15,500 s. For inland zones, the average training time is 3900 s and the prediction time is 16,400 s, while for water zones, the processing time is shorter with an average training time of 3100 s and prediction time of 15,000 s.

These results show that the *GHS-S2Net*-based multi-modeling approach scales seamlessly in a distributed multi-GPU platform. For the processing at a global scale, our main constraint was the limited amount of concurrently available GPUs: we employed 6 GPU modules for the training phase and 2 modules for the prediction phase that were available at the time of deployment. Despite these limitations, we managed to scale up the *GHS-S2Net*-based multi-modeling approach and achieved to process a dataset having global coverage at 10 m spatial resolution thanks to: (i) an efficient partitioning of the processing per UTM grid zone, (2) the two-stage training approach with a subsampling of non-built-up patches within the selected tiles containing training samples, and (3) the optimal size of input data (i.e., 100 × 100 km tiles) used for both the training and prediction. Increased GPU capacities and activation of early stopping during the training in order to reduce the number of iterations (epochs) when the loss function stops improving, can significantly reduce both the training and the prediction time of the GHS-S2Net model.

## 3.3 Qualitative assessment of the models predictions

The results of the *GHS-S2Net* implementation on the Sentinel-2 global mosaic were assessed visually. Compared to the training sets, the results of built-up detection showed a significant reduction in both commission and omission errors and other artifacts that were observed in the training sets (see Sect. 2.2). In addition, *GHS-S2Net* resulted in a refined mapping of built-up areas and open spaces within urban areas and most importantly the detection of new settlements, never annotated so far in the training sets or identified in any other global scale dataset. Figure 9 illustrates some examples of each type of improvement obtained with the *GHS-S2Net* models. Figures 9, 10 and 11 show, for selected cities, the enhanced built-up areas detection, represented in the form of continuous-range outputs (probability), in comparison with the best available training sets. The most notable improvements relate to the detection of built-up areas which are omitted from the training sets, under the assumption that the initial purpose of these datasets was to map completely the contiguous areas they cover. These omissions are either due to lack of data or to flaws and gaps in the training sets themselves given that they were all extracted through automatic classification of satellite imagery. In the case of FB_HRS (Fig. 9a: 7.34 Latitude, 3.90 Longitude), the most critical omissions were systematically observed in dense built-up areas (often corresponding to urban cores), while in ESM_BU (Fig. 9b: 51.44 Latitude, −0.97 Longitude), the omissions were essentially due to lack of input satellite data in some countries (mainly United Kingdom and Ireland). In the case of MS_BFP (Fig. 9c: 43.11 Latitude, −79.05 Longitude), most of the omissions concerned large industrial buildings but several small buildings were also not detected in this training data. For GHSL_BU (Fig. 9d: 30.51 Latitude, 120.67 Longitude), underdetections were mainly observed in rural areas and in particular in small scattered settlements due to the size of the built-up structures which were difficult to be captured due to the sensor's spatial resolution.

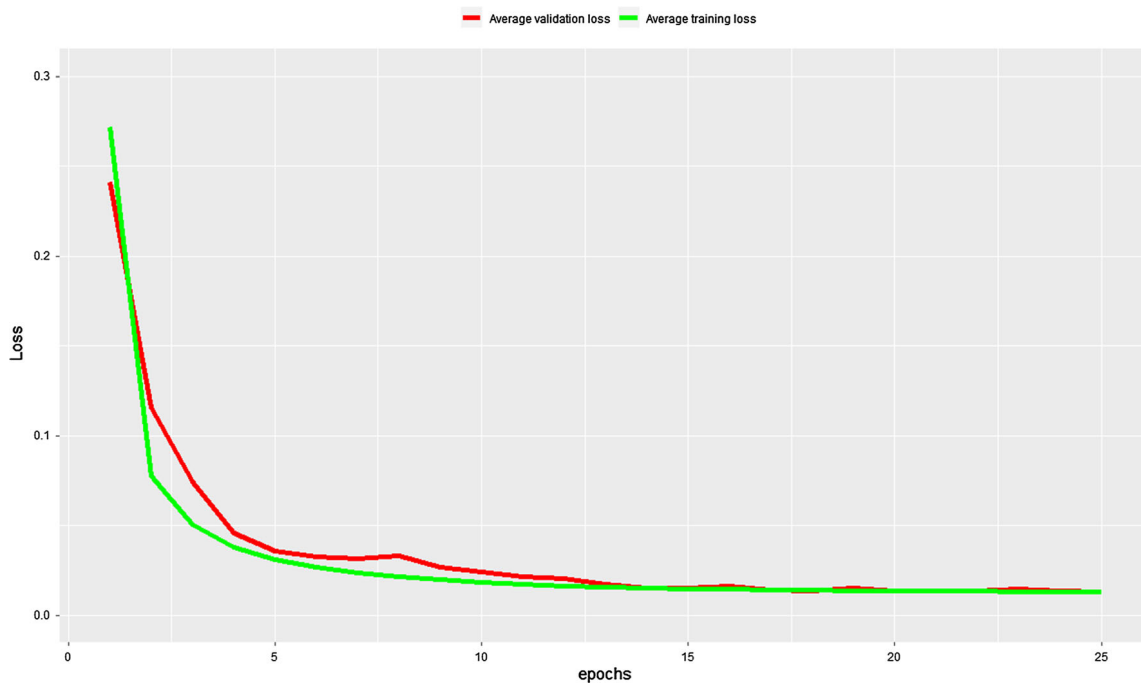

**Fig. 7** Average loss curves produced by 485 GHS-S2Net models during their training and validation, where each model corresponds to a different UTM grid zone (color figure online)

Figure 10 is another example highlighting the capacity of the *GHS-S2Net* in reducing significantly commission problems observed in the training sets that were fed to the models. In the case of MS_BFP, overdetections were mainly observed in mountainous areas with bare rocks or in agricultural areas with bare fields (Fig. 10a: 33.25 Latitude, −90.62 Longitude). In the case of ESM_BU, overdetections were frequently identified in sand dunes (Fig. 10b: Latitude 43.36, 16.65 Longitude) and rocky beaches, bright bare soils and riverbeds.

The visual comparison of the results of the *GHS-S2Net* probabilistic output against the best available training sets provides a clear evidence of the refined built-up areas detection from the Sentinel-2 image composite. Figure 11 is an example of such enhanced capabilities covering the city of Sassari (Italy). It compares the ESM_BU training set derived from VHR satellite data at a spatial resolution of 2 m to the results obtained by the *GHS-S2Net* trained with ESM_BU. These results illustrate the unprecedented performance of *GHS-S2Net* for pixel-wise classification of 10 m Sentinel-2 data and for detecting urban structures in complex urban environments. Not only the classification of built-up areas is more refined, despite the coarser spatial resolution of Sentinel-2 data (10 m) in comparison with the VHR imagery used for producing ESM_BU (2 m)

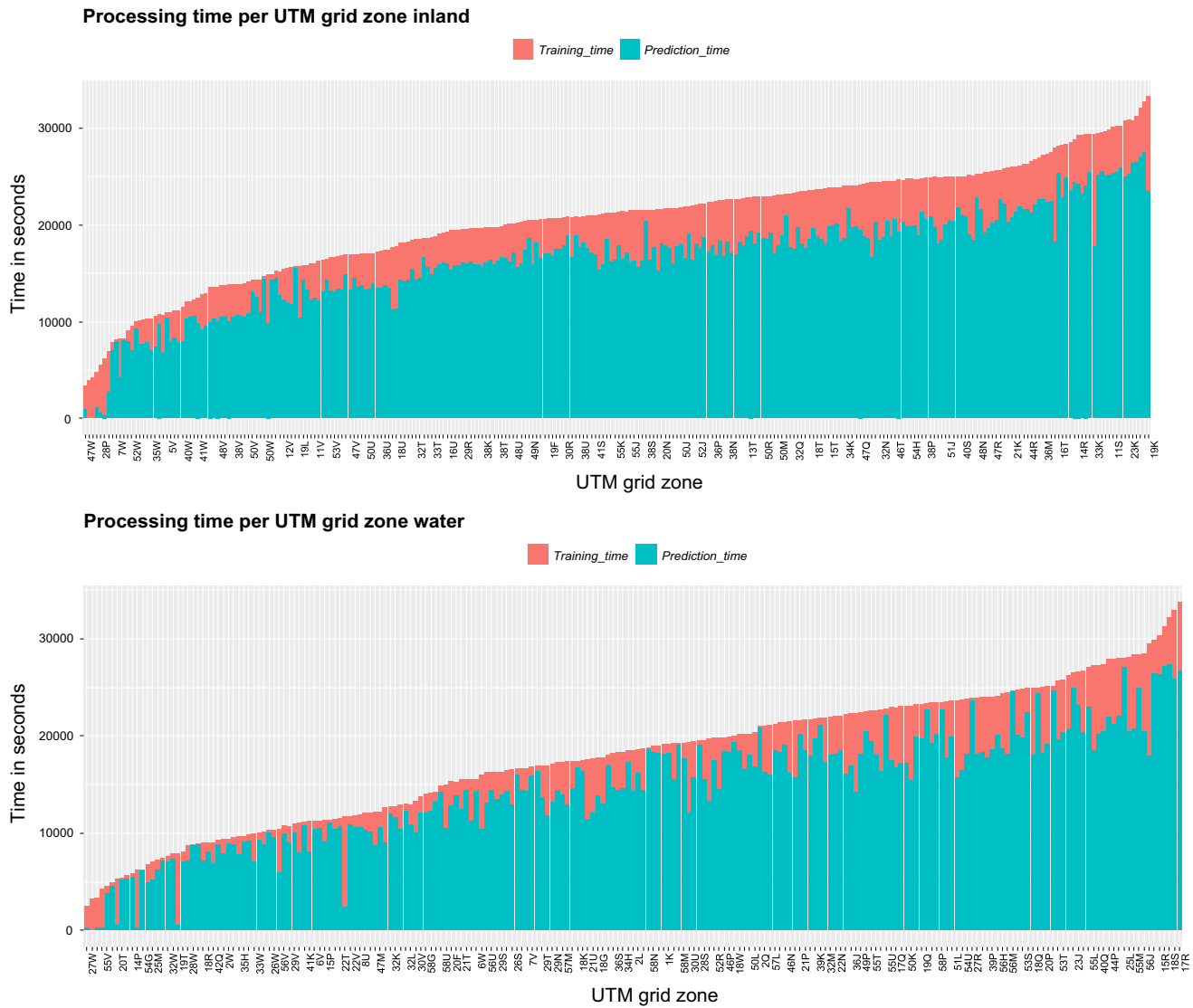

**Fig. 8** Training and prediction runtime per UTM grid zone. The upper figure refers to inland dominated grid zones and the bottom figure to the water dominated grid zones. Islands fall within water dominated UTM grid zones

(Fig. 11b), but it is almost possible to identify single buildings as well as open spaces in the urban layout. Besides, the probabilistic output seems to be highly related to the patterns of built-up areas suggesting that *GHS-S2Net* may be a proxy measure for building densities.

These examples provide experimental findings that support the *GHS-S2Net* model generalization capacity, which was already evidenced during the training phase (3.1.2). With a relatively small number of parameters (1,447,042 trainable parameters) and a very large number of samples (511,502,073 total number of built-up patches—See Supplementary material R1 for training samples per UTM zone), the model proved to be robust to noise or missing data with respect to the training sets, while effectively capturing the essential patterns and salient features, resulting in precise mapping of built-up areas.

### 3.4 Validation of the model predictions and assessment of generalization performance

Two approaches were implemented for the validation of the *GHS-S2Net* output that are based on comparison with independent cartographic data of building footprints, not employed for the training of the models:

- Continuous assessment: by testing the *GHS-S2Net* output as predictor of the built-up densities at the spatial resolution of 10 m through least-square linear regression;

**Fig. 9** Example predictions of GHS-S2Net in the form of probabilities of built-up areas. One example is given per each training set to demonstrate the benefit of the model output compared to the input best available training sets (**a** FB_HRS, **b** ESM_BU, **c** MS_BFP, **d** GHSL_BU)— Google satellite imagery is used in the background

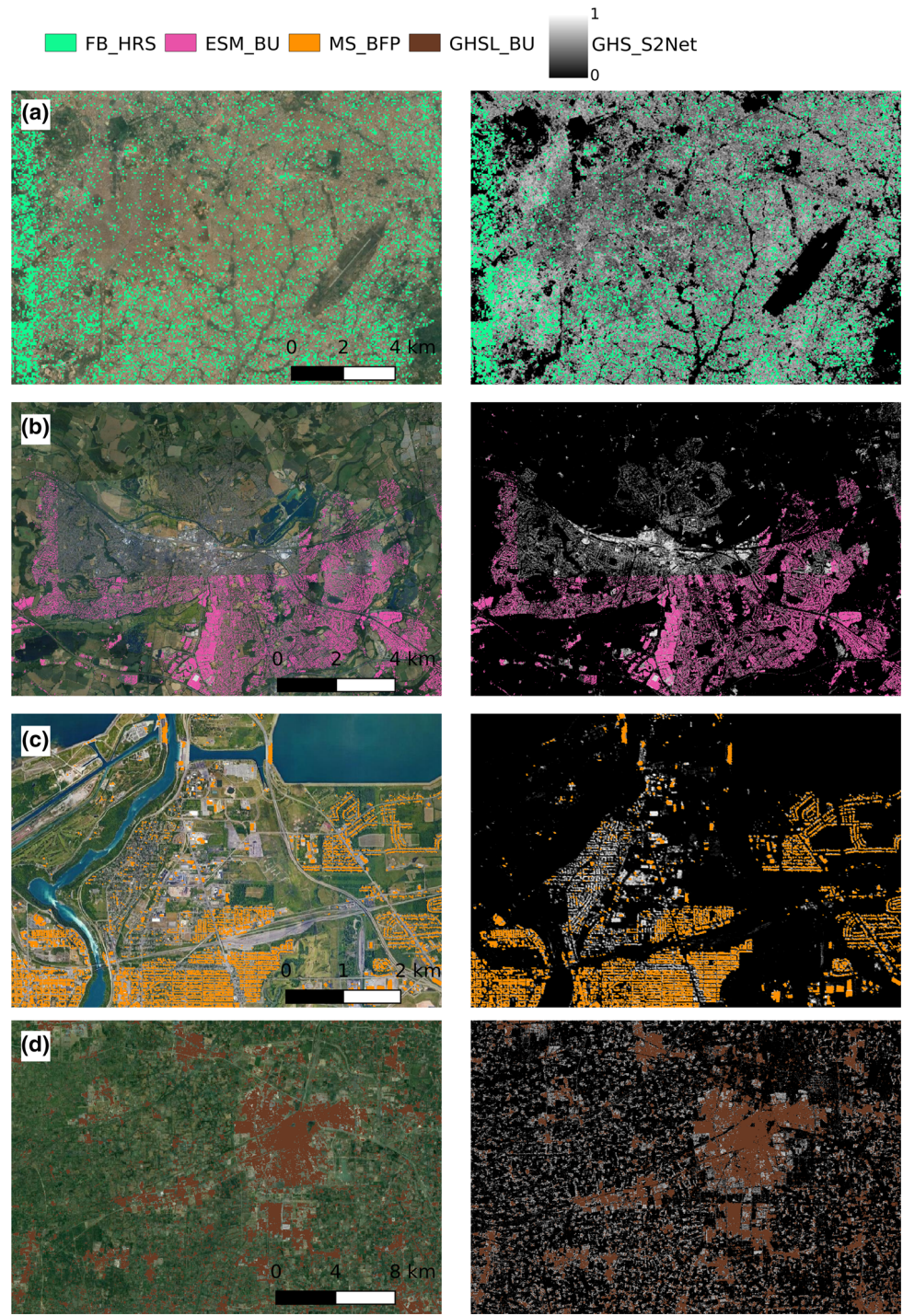

- Binary assessment: by evaluating the contingency table between the binarized outputs of *GHS-S2net* after the application of a probability cut-off value, and the binarized reference data used as a "ground-truth."

For the validation of pixel-wise predictions, a reference spatial database including single building delineation derived from digital cartography at a nominal scale of 1:10,000 was developed. The suitability of this database for the global scale validation of built-up products derived from remote sensing data has been previously evaluated in Corbane et al., 2019 [5]. The reference database consists of more than 40 million individual building polygons selected from 277 different areas of interest (AOI) around the globe. These are mostly local administrative units covering specific cities or full counties (for the United States of America) and spread across different continents. While not covering all the combinations of geographical, environmental, and cultural conditions that are determinant factors of the settlement patterns, the reference data spread across various landscapes. The reference years for the collected reference data range between 2012 and 2018 with the latter being the most frequent year of update. This makes the reference database suitable for the validation of the results derived from the Sentinel-2 pixel based image composite produced for the reference year 2018. The building footprints span over the whole spectrum of low-density and high-density human settlement patterns, representing typical rural, suburban and urban spatial patterns (see supplementary material R2 for more information on the spatial distribution and characteristics of the reference dataset). In order to support the accuracy assessment exercise, the reference data collected in vector format were converted into binary raster layers indicating the presence/absence of built-up areas. The rasterization of the vector cartographic data was performed at a spatial resolution of 10 m corresponding to the spatial resolution of the Sentinel-2 image composite and the outputs of the *GHS-S2Net* model.

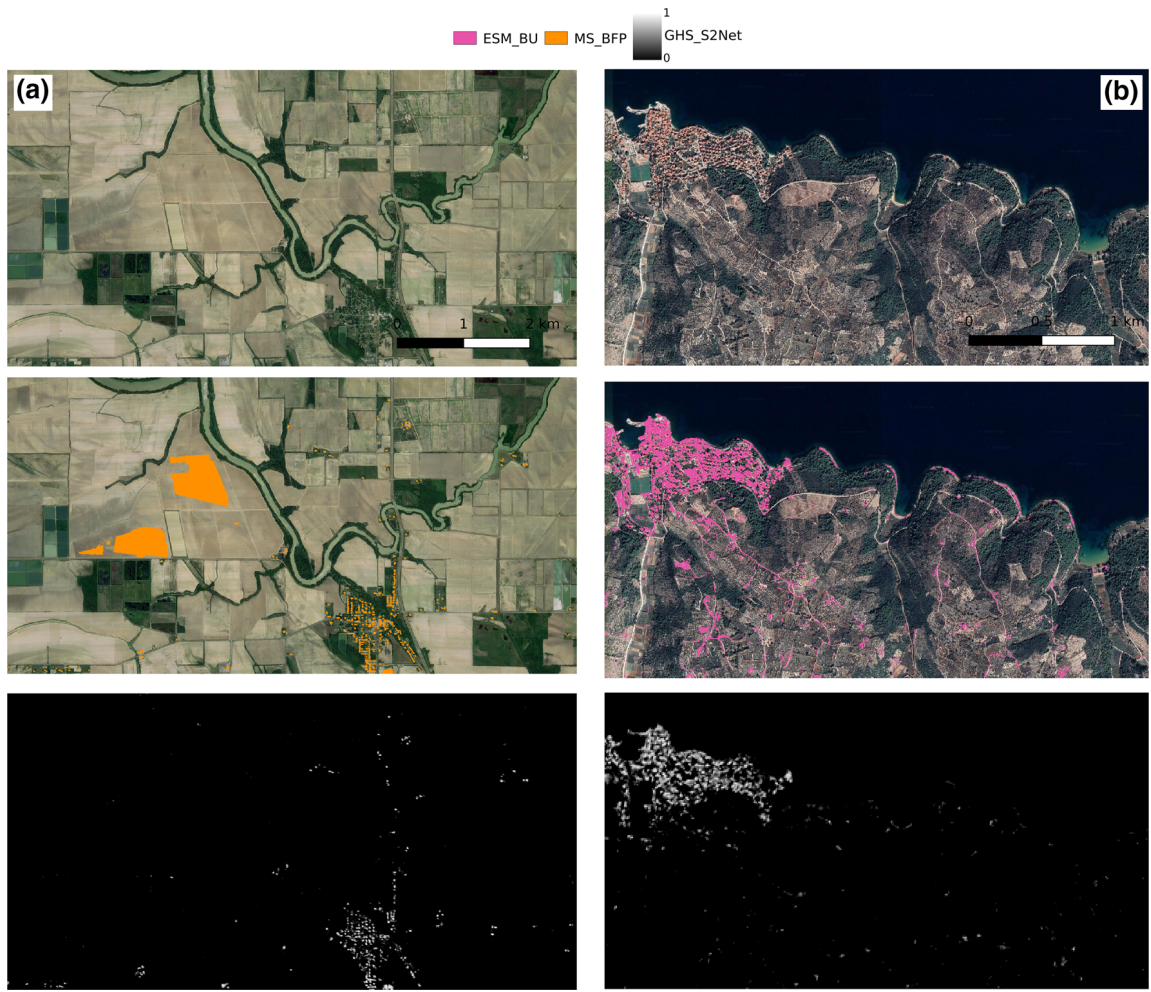

**Fig. 10** Examples of reduction and mitigation of commission errors. The figures show a comparison between the probability of built-up areas obtained from the GHS-S2Net and the input training sets based on **a** MS_BFP in Northern America and **b** ESM_BU in Europe—Google satellite imagery is used in the background

### 3.4.1 Continuous assessment: validation of the model output as predictor of built-up densities

For analyzing the performance of the *GHS-S2Net* model as a predictor of the densities of built-up areas, we perform a regression analysis between the probability of built-up areas given by the model as response and the reference built-up surface densities as derived from the database of building footprints for the 277 different areas of interest. The knowledge of the systematic bias and gain parameters of the automatically classified built-up areas allows us to gain insights into the capacity of the *GHS-S2Net* model in capturing the patterns and densities of built-up areas and to identify a suitable threshold for the binarization of the output probabilities for the subsequent accuracy assessment step.

**Fig. 11** Example of refined built-up areas detection in the city of Sassari with different types and densities of buildings. **a** Extract from VHR Google imagery, **b** ESM_BU training set derived from Copernicus VHR_2015 and **c** output of GHS-S2Net representing probabilities in built-up areas—Google satellite imagery is used in the background

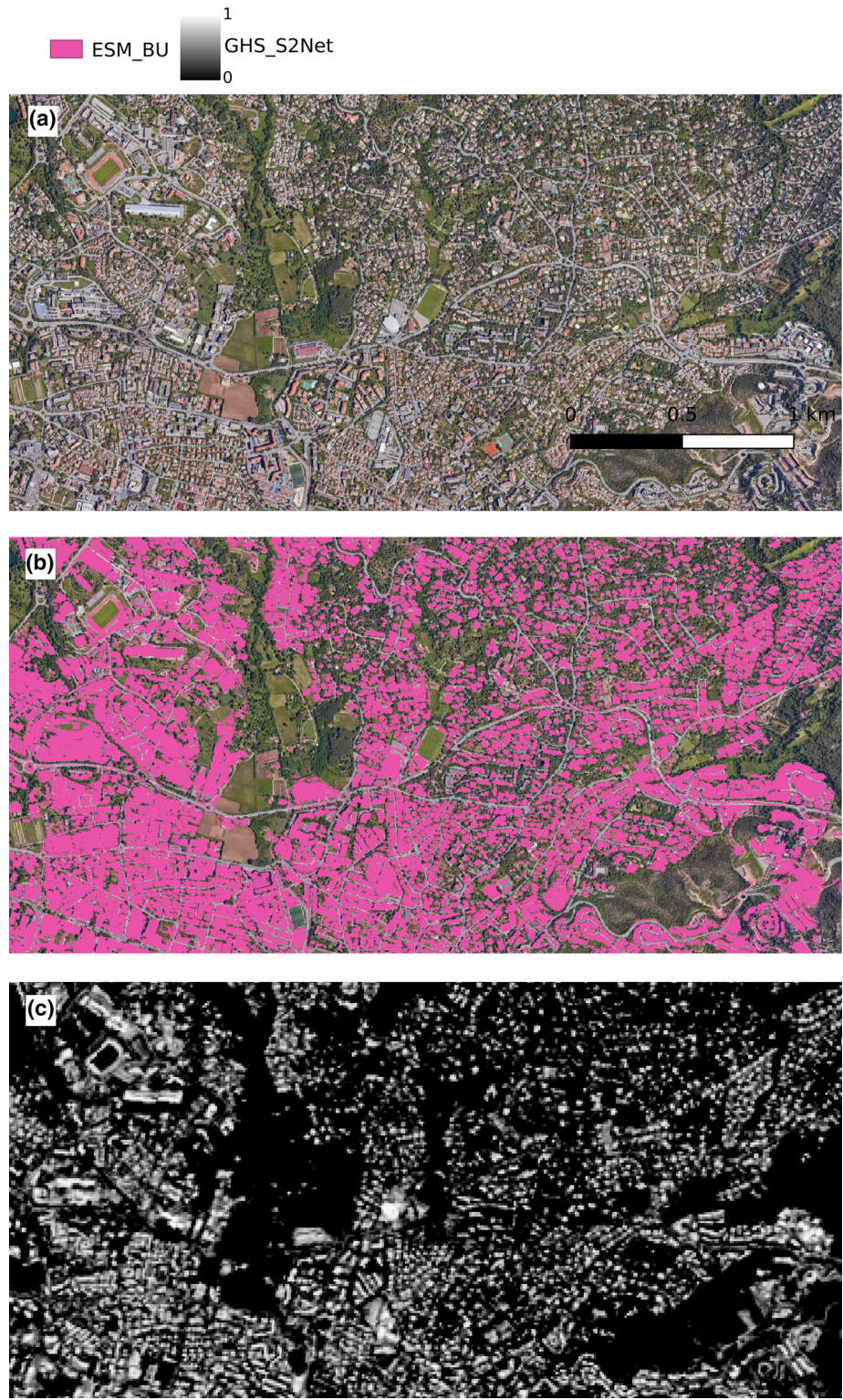

The strength of the linear relation between the automatically generated built-up probabilities and the reference data is assessed through the Pearson correlation coefficient

(r). The gain factor (slope) allows the user to model, retrofit and compare the results obtained from the *GHS-S2Net* model for the different AOIs. In addition, the slope of the regression is an indicator of the optimal threshold for

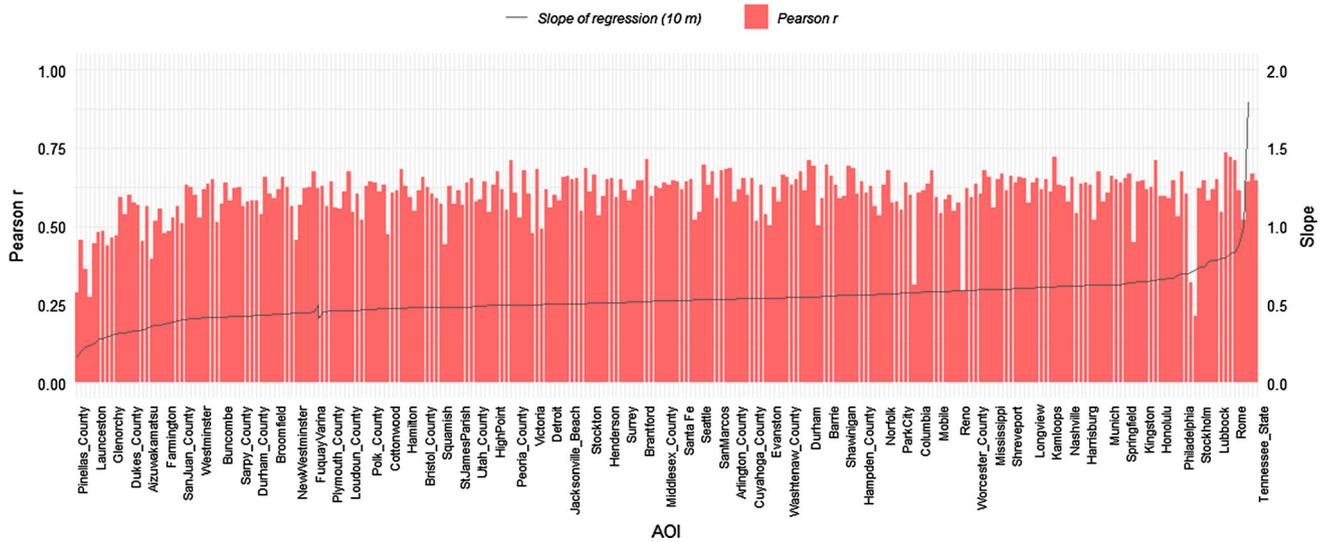

**Fig. 12** Results of the regression analysis between the output probabilities of built-up areas and built-up densities at 10 m resolution. The results represented here by the correlation coefficient (r) and the slope of regression are shown for the 277 AOI

translating the built-up probabilities to binary values for the pixel-based accuracy assessment.

The results of the regression analysis at 10 m for all AOI sites showed an average correlation coefficient r of 0.67 and an average slope of 0.52 (Fig. 12).

The average correlation coefficient shows that the output probabilities from *GHS-S2Net* models are capable of capturing around 67% of the structural variability in built-up areas. The lowest correlation coefficients were observed for AOIs covering complete counties in the United States where there are a lot of building sizes below 100 m$^2$ (which is the limit of the detectability of the Sentinel-2 sensor) and the built-up density is very low, less than 0.5%. This is for instance the case of the Matanuska-Susitna Borough AOI which is a borough located in the state of Alaska, covering an area 9492.46 km$^2$ with a built-up density of 0.1% and an average size of buildings of 140 m$^2$ (Supplementary material R2). The output probabilities of the *GHS-S2Net* models seem to better capture building densities in urban areas and high density AOIs where the correlation coefficients were greater than 0.6. This is the case for example of the AOI covering San Francisco city with an area of 194 km$^2$ and a building density of 26.4%.

It is also worth noting that the gain factor (slope) translating the built-up probabilities as derived from Sentinel-2 data to built-up surface densities as derived from the reference cartographic data is almost constant. The slope has an average of 0.2 in low density AOIs, in particular those covering full counties in the United States (e.g., San Juan County). In high-density AOIs covering cities, the slope (bias) is higher (e.g., city of Rome where the slope is close to 0.8) with an average around 0.54.

According to these findings, it is not straightforward to define one general-purpose threshold to binarize the output of the *GHS-S2Net* models into two classes "built-up" and "non-built-up." A threshold of 0.2 would then be good compromise targeting large areas including scattered settlement patterns, in particular rural areas, while a more conservative threshold of 0.5 would be more suitable for areas largely dominated by high-density built-up areas (i.e. city centers). Following this finding, both thresholds were applied to the outputs of the *GHS-S2Net* models for assessing the quality of the classifications following a pixel-wise accuracy method.

### 3.4.2 Binary accuracy assessment

The thresholds 0.2 and 0.5 identified in the previous regression analysis were used to binarize the probabilistic output as required by the pixel-wise binary accuracy assessment at the spatial resolution of the sensor. Standard accuracy and error metrics derived from the confusion matrix were calculated for the binary results obtained with the two thresholds. Given the lack of a single universally accepted measure of agreement, we use a combination of two main performance metrics to give a complete picture of the performance of the *GHS-S2Net* models: the balanced accuracy and the Kappa coefficient that were introduced to the remote sensing community and recommended by Congalton, 2011 [87]. The Balanced Accuracy and Kappa are measures of classification accuracy, the former providing information about the rate of correctly classified pixels in an unbalanced setting where non-built-up pixels are predominant compared to built-up pixels. The latter compensates for random chance in the pixels assignment.

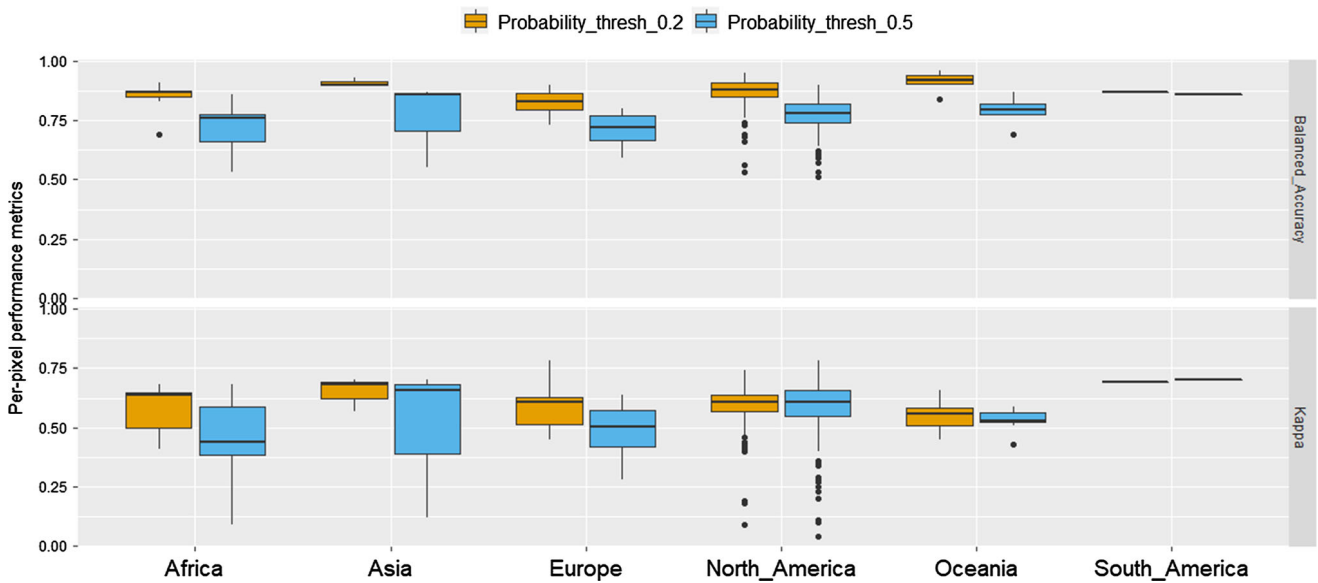

**Fig. 13** Per-continent, box plots of the performance metrics (Balanced Accuracy and Kappa) of the two binary classification outputs obtained by applying 0.2 and 0.5 thresholds to the probability outputs

The results of the per-pixel accuracy assessment with the two binary outputs are summarized in Fig. 13 and disaggregated per continent. The figure shows the average and standard deviations of the Balanced Accuracy and Kappa coefficients per binary output and per continent. The 277 AOI were grouped by continent to evidence major improvements, especially in areas where previous global products failed to produce satisfactory results. Overall, both binary classifications produce good results with an average Balanced Accuracy greater than 0.7 and an average Kappa greater than 0.5. However, when compared to the binary outputs derived with the 0.5 probability threshold, the classification with a less conservative threshold of 0.2 produces better agreement with the reference data, consistently for all continents. The best results in the least conservative classification outputs (threshold of 0.2) were obtained in Oceania an Asia with an average Balanced Accuracy of 0.91, followed by North America and Africa where the mean Balanced Accuracies were equal to 0.86 and 0.85, respectively.

The results of the per-pixel accuracy assessment, in particular those obtained by applying a low threshold to the probability outputs, constitute a strong evidence of the modeling power of the *GHS-S2Net* and the reliability of the outputs. They are also a confirmation of the merit of the new classification framework for identifying settlements in challenging landscapes such as in Africa and Asia. They also suggest that for the generation of a global binary classification from the probabilistic output of the models, a low probability threshold is recommended, in particular if the purpose is to capture all the scattered settlements in rural landscapes such as in Africa. In this particular context, the binary outputs obtained with a threshold of 0.2 outperform significantly those derived from the conservative threshold.

## 3.5 Comparison between the results of close range and far range transfer learning

When computing the *GHS-S2Net* predictions at the global scale, the majority of the UTM grid zones and in particular the $100 \times 100$ km$^2$ tiles were processed with the close range transfer learning. However, to allay the scarcity and quality issues in the training dataset, 28 UTM grid zones were classified according to the far range transfer learning and the outputs were compared to those obtained by the direct close range transfer learning. Figure 14a illustrates the differences between close range (middle figure) and far range transfer learning (bottom figure) in areas suffering from the lack of training samples (e.g., in Ethiopia). It shows the capacity of the far range transfer learning in discovering undetected built-up features in UTM grid zone 37P, on the basis of the parameters of the model trained in the neighboring UTM grid zone 37 M. In such a situation, the close range transfer learning was less effective in identifying those scattered settlements due to insufficient training samples in the UTM grid zone 37P.

Figure 14b shows another example with respect to the city of Moscow, showing the added-value of the far range transfer learning in areas where only the GHS_BU low-resolution training data were available (UTM grid zone 37U). The example highlights the generalization capacity of the *GHS-S2Net* trained on a UTM grid zone where detailed training samples are available (e.g., in UTM grid

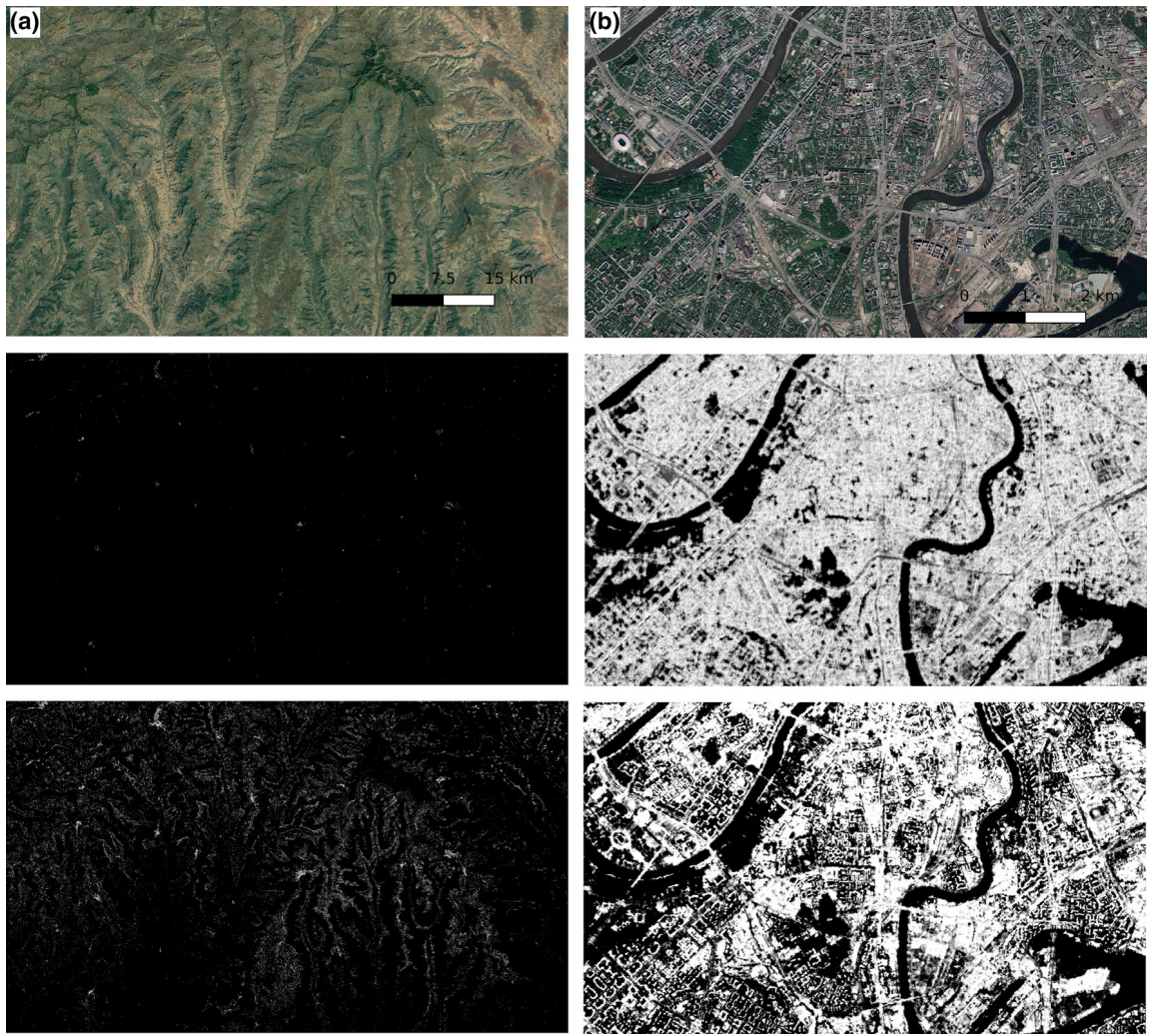

**Fig. 14** Comparative results of the close range (middle figures) and far range transfer learning (bottom figures) in **a** Ethiopia, **b** Moscow (Russia)—Google satellite imagery is used in the background

zone 34U) and then applied to the nearby zone. The generalization capacity of the model here is reflected in: (i) reproducing fine-scale settlement structures in dense built-up areas, (ii) reducing overdetections of roads and other impervious features and (iii) enhancing the sharp delineation of buildings and open spaces in the built-up areas.

Moscow is one of the cities where detailed building footprints were available in the reference database used in the validation exercise. The availability of "ground-truth" data enabled to conduct a quantitative binary accuracy assessment of the results of far range transfer learning in comparison with those obtained with the close range transfer learning. The results are illustrated in Table 2 for the binary outputs with cut-off values of 0.2 and 0.5. They show higher overall and balanced accuracy values resulting from the application of far range transfer leaning. These results are an additional evidence of the enhanced mapping capabilities of a well-designed far range transfer learning approach deployed in this work.

The encouraging results were determinant for expanding the application of far range transfer learning which was finally implemented on a total of 28 UTM grid zones. The selection of source and target UTM grid zones was mainly driven by spatial adjacency or similarities in the landscape and in the type of built-up areas.

**Table 2** Summary Results of binary accuracy assessment of the close range and far transfer learning in the city of Moscow based on detailed building footprints

| | Overall accuracy | | Balanced accuracy | |
|---|---|---|---|---|
| | 0.2 cut-off | 0.5 cut-off | 0.2 cut-off | 0.5 cut-off |
| Close range transfer learning | 0.61 | 0.67 | 0.75 | 0.76 |
| Far range transfer learning | 0.77 | 0.83 | 0.81 | 0.78 |

# 4 Discussion and future work

In this paper, we presented a novel end-to-end framework for large-scale pixel-wise classification of built-up areas from high-resolution satellite imagery. The developed multi-model approach designated by *GHS-S2Net* builds on a relatively simple CNN architecture. The implementation of the models on a global cloud-free Sentinel-2 image composite provides the most detailed and complete map reporting about built-up areas in the form of probability outputs (i.e., probability of a pixel to belong to the class 'built-up'). The results confirm the high generalization capacity of the model and its ability to not only detect new built-up areas in difficult landscapes (i.e., in Africa and Asia) without site-specific training sets, but also its potential to mitigate commission errors in the best available training sets reporting about built-up areas across the globe.

The implementation of the developed framework for large classification of human settlements was achieved thanks to three main building blocks:

- The multi-neuro modeling methodology, which follows the UTM grid zones schema and the systematic sampling within each UTM grid zone. This approach of training multiple lightweight models at global scale allows decomposing the optimization phase into smaller tasks, which are then solved in parallel. The adopted sampling approach meets the three following criteria: class balance, diversity, and representativeness. It shows to be suitable for an optimal learning of the models at a global scale without compromising performance;

- The transfer learning includes both the close range and the far range transfer learning. Both approaches benefit from parameter-based transfer methods where the optimal parameters found in the source domain classifier are used for the target domain. The novelty of the approach implemented in the paper was the use of the close range transfer learning within the same UTM grid zone in a way to alleviate the computational burden and avoid overfitting issues. The far range transfer learning leverages the optimal parameters found when training the models with detailed and high-quality training sets in a given UTM grid zone and then applying them to neighboring zones subject to training data scarcity. The far range transfer learning allowed allaying the scarcity and quality issues in the training sets while achieving outstanding performance in the reduction of commission and omission errors in the best available training data and in the refinement of built-up areas detection;

- The deployment of the high-throughput processing, including data preparation, learning and inference on the multi-petabyte scale JEODPP platform. The big data multi GPU platform enables: (i) the efficient storage of the large volume of input satellite data (15 TB) and the output (1.5 TB) maps encoded in 16 bits, (ii) the parallel training of the models on an heterogeneous cluster of GPUs, and the (iii) optimal load balance in terms of data retrieval and processing from and to the distributed system due to the efficient co-location of the data with the processing units.

The validation of the results with an independent reference dataset of building footprints covering 277 sites across the world establishes the reliability of the built-up layer produced by the *GHS-S2Net* approach and the model robustness against both the variable conditions in the satellite imagery and the heterogeneity in the landscapes and built-up characteristics. The most noticeable achievement is the capacity of the model to classify built-up areas in remote areas (e.g., in Africa and in Asia), reported in none of the global products (i.e., GUF, WSF, FROM GLC10). Further research is currently being deployed to quantitatively assess the improvements gained with the *GHS-S2Net* compared to other global products reporting on built-up areas.

Another significant result is the strong relationship between the output probabilities and the building densities suggesting that the model outputs can be used as proxy measures for building densities without additional calibration or modeling.

Despite the unprecedented results obtained by the proposed approach on an extremely challenging dataset in terms of spatial coverage, resolution and spectral variability, some challenges need to be considered, especially if the aim is to regularly update the built-up layer for continuous monitoring of human settlements with Copernicus Sentinel-2 data. The challenges pertain to methodological choices when designing the model and during its scaling to the classification of the global composite:

- The choice of patch size: in general, assessment of CNN accuracy indicates that using larger patch sizes yields higher accuracies because the network is able to learn more contextual features. In the case of the Sentinel-2 pixel-based classification, the experiments performed by [71] on Sentinel-2 data showed that larger patch sizes (e.g., 15 × 15) did not yet yield significant improvement in the model accuracy. In this work, we tested a 10 × 10 patch size resulting in a deeper network topology, yet the loss function did not improve during the training phase whereas the prediction accuracy worsened.

- The far range transfer learning: the strategy for implementing the far range transfer learning was based on criteria related to spatial adjacency of UTM grid zones or similarities in the landscape and in the type of built-up areas. The potential of this approach for mitigating problems in the training data and for deriving fine-grained classification outputs was clearly demonstrated in the classification results. Nevertheless, the added-value of this approach was not fully exploited in the context of this work. Additional work should focus on the analysis of spatial patterns of landscape features and typologies of built-up areas and their influence on the outputs of the classification with *GHS-S2Net*. The ultimate goal is to unveil the underlying rules and associations for designing a more systematic approach to identify the source and the target UTM grid zones candidate for the far range transfer learning.

- The variable quality of the training data: despite their outstanding learning capability, the lack of accurate training data might limit the applicability of CNN models in realistic remote-sensing contexts [88]. For our global scale application, the strategy was to collect the best publicly available training data and reporting about built-up areas. The higher the spatial resolution of the training data, the more detailed is the output of the classification. Ideally, the spatial resolution of the input training data should be equal or better to that of the input Sentinel-2 imagery. As described in Sect. 2.2, the reference data sources have variable spatial resolutions. In addition, the trustworthiness of samples is highly variable across the different sources but also within same reference data source. The lack of consistency in the training data produces outputs with variable qualities depending on the input data used for training the models. This was reflected by the results of the validation when disaggregated per continent. One approach to deal with imperfect training data was to use the far range transfer learning. However, this approach has a limited applicability at global scale since it supposes that the target UTM grid zones have similar characteristics (in terms of landscape and types of built-up areas) with the source zones. Another approach is to use a two-step training approach in which the models are first initialized by using a large amount of possibly inaccurate reference data, and then refined on a small amount of accurately labeled data, similarly to the method developed in Maggiori et al. [88]. In the context of our large-scale classification, it is perfectly reasonable to use the output produced by the *GHS-S2Net* to train a new model. The use of high quality and consistent outputs produced for the reference year 2018 by the application of the *GHS-S2Net* model at global scale is a key for frequent updates of built-up layers from Sentinel-2 Copernicus data and for continuous monitoring of built-up areas.

The outputs of this work are an additional demonstration of the far-reaching opportunities of using DL in remote sensing applications. They set the path to a new wave of promising research to tackle unprecedented large-scale challenges in urban related applications building on remote sensing data. Further research is currently being carried out to explore the potentials of using DL and EO to assess changes in built-up areas and to generate consistent time series reporting about the dynamics of human settlements at the European and global scales. Characterization of the urban environment is another research field in which DL and EO are bringing new advances and a concrete opportunity to develop models for describing the buildings' typology and the morphology of the cities, hence addressing global interrelated challenges of urbanization and climate change.

## Compliance with ethical standards